\documentclass[a4paper,12pt]{article}
\pdfoutput=1
\usepackage{graphicx}
\usepackage{mathtools}
\usepackage{amssymb}
\usepackage{amsfonts}
\usepackage[footnotesize]{caption}
\usepackage[font=scriptsize]{subcaption}
\usepackage{color}
\usepackage{braket}
\usepackage{cite}
\usepackage{hyperref}
\usepackage{url}
\usepackage{multirow}
\usepackage{relsize}
\usepackage{fullpage}
\usepackage{makecell}
\usepackage{blkarray}
\usepackage{fullpage}

\newcommand{\newc}{\newcommand}
\newc{\be}{\begin{equation}}
\newc{\ee}{\end{equation}}
\newc{\bea}{\begin{eqnarray}}
\newc{\eea}{\end{eqnarray}}
\newc{\simlt}{~\mbox{\smaller\(\lesssim\)}~}
\newc{\simgt}{~\mbox{\smaller\(\gtrsim\)}~}

\newcommand{\pmatr}[1]{\begin{pmatrix} #1 \end{pmatrix}}
\def\RK{R_{K^{\ast}}}

\begin{document}

\begin{titlepage}
\begin{center}
{\bf\Large
\boldmath{
$R_{K^{(*)}}$ and the origin of Yukawa couplings
}
} \\[12mm]
Stephen~F.~King$^{\star}$%
\footnote{E-mail: \texttt{king@soton.ac.uk}}
\\[-2mm]
\end{center}
\vspace*{0.50cm}
\centerline{$^{\star}$ \it
Department of Physics and Astronomy, University of Southampton,}
\centerline{\it
SO17 1BJ Southampton, United Kingdom }
\vspace*{1.20cm}

\begin{abstract}
{\noindent
We explore the possibility that the semi-leptonic $B$ decay ratios $R_{K^{(*)}}$ which violate $\mu - e$ universality
are related to the origin of the fermion Yukawa couplings in the Standard Model. 
Some time ago, a vector-like fourth family (without a $Z'$)
was used to generate fermion mass hierarchies and mixing patterns without introducing any family symmetry.
Recently the idea of inducing flavourful $Z'$ couplings via mixing with a vector-like fourth family which carries gauged $U(1)'$ charges
has been proposed as a simple way of producing controlled flavour universality violation while elegantly cancelling anomalies. 
We show that the fusion of these two ideas provides a nice connection between $R_{K^{(*)}}$ and the origin of Yukawa couplings in the quark sector. However the lepton sector requires some tuning of Yukawa couplings to obtain 
the desired coupling of $Z'$ to muons.
}
\end{abstract}
\end{titlepage}

\section{Introduction}

Recently, the phenomenological motivation for considering non-universal $Z'$ models has increased due to 
mounting evidence for semi-leptonic $B$ decays which violate $\mu - e$ universality at rates which exceed those predicted by the SM
\cite{Descotes-Genon:2013wba,Altmannshofer:2013foa,Ghosh:2014awa}.
In particular, the LHCb Collaboration and other experiments have reported a number of anomalies in $B\rightarrow K^{(*)}l^+l^-$
decays such as the $R_K$ \cite{Aaij:2014ora} and $R_{K^*}$ \cite{Bifani} ratios of $\mu^+ \mu^-$ to $e^+ e^-$ final states, 
which are observed to be about $70\%$ of their expected values with a $4\sigma$ deviation from the SM,
and the $P'_5$ angular variable,
not to mention the $B\rightarrow \phi \mu^+ \mu^-$ mass distribution in $m_{\mu^+ \mu^-}$.

Following the measurement of $R_{K^*}$ \cite{Bifani}, a number of phenomenological analyses of these data, 
see e.g. \cite{Hiller:2017bzc,Ciuchini:2017mik,Geng:2017svp,Capdevila:2017bsm,Ghosh:2017ber,Bardhan:2017xcc,
Glashow:2014iga,DAmico:2017mtc,Descotes-Genon:2015uva,Calibbi:2015kma}
favour a new physics operator of the form 
$\bar b_L\gamma^{\mu} s_L \, \bar \mu_L \gamma_{\mu} \mu_L$,
or of the form, $\bar b_L\gamma^{\mu} s_L \, \bar \mu \gamma_{\mu} \mu$,
each with a coefficient $\Lambda^{-2}$ where $\Lambda \sim 31.5$ TeV,
or some linear combination of these two operators. 
For example, in a flavourful $Z'$ model, the new physics operator will arise from tree-level $Z'$ exchange,
where the $Z'$ must dominantly couple to $\mu \mu $ over $e e $, 
and must also have the quark flavour changing coupling $b_L s_L$ which must dominate over $b_R s_R$.
There is a large and growing body of literature on non-universal $Z'$ models as applied to explaining
these anomalies \cite{large}, with more recent papers
following the $\RK$ measurement in~\cite{recent}.
One of the challenges facing such models is the requirement that they be anomaly-free.

In the present paper we will investigate 
the possible connection between the experimental signal for new physics in 
$R_{K^*}$ and the origin of fermion Yukawa couplings.
Remarkably, despite the huge recent literature on $R_{K^*}$,
relatively few papers are concerned with its possible connection with Yukawa couplings \cite{Aloni:2017ixa}.
Here we shall take the point of view that the observation of $R_{K^*}$ 
is the first hint that the origin of Yukawa couplings is just round the corner close to the TeV scale. 
We shall focus on $Z'$ models where the physics responsible for the $Z'$ mass and couplings is also responsible for 
generating the effective Yukawa couplings. This is the first time that such a connection has been considered in the literature to our knowledge. 

We consider a scenario in which the Standard Model (SM) at the electroweak scale is an effective theory
resulting from some theory at some higher scale(s) which may be as low as the TeV scale.
We require that all fermion Yukawa couplings must result from higher dimension operators, so that the effective Yukawa couplings of the SM can be expressed in terms of the left-handed fermion electroweak doublets,
$\psi_i=L_i,Q_i$, where $i=1,2,3$, and the CP conjugated right-handed electroweak singlets, 
$\psi^c_j=u^c_j,d^c_j,e^c_j,\nu^c_j$,
\be
{\cal L}^{Yuk}_{eff}=
 \left(\frac{\langle\phi_i \rangle}{\Lambda^{\psi}_{i,n} }\right)^n
\left(\frac{\langle \phi_j \rangle}{\Lambda^{\psi^c}_{j,m}}\right)^m
H \psi_i \psi^c_j ,
\label{1}
\ee
plus $H.c.$, summed over fields, families and powers of $n,m$.
Eq.\ref{1} involves new SM singlet fields $\phi_i$
which develop VEVs, leading to effective Yukawa couplings suppressed by powers of 
$\langle \phi_i \rangle /\Lambda$.
Our scenario also involves a massive $Z'$
under which the three SM families $\psi_i$ have zero charge, and 
which only couples to it via the same singlet fields $\phi_i$ which have non-zero charge
under the associated $U(1)'$ gauge group,
\be
{\cal L}^{Z'}_{eff}=
 \left(\frac{\langle \phi_i\rangle }{\Lambda'^{\psi}_{i,n} }\right)^n
\left(\frac{\langle \phi_j \rangle}{\Lambda'^{\psi}_{j,m}}\right)^m
g'Z'_{\mu}\psi_i^{\dagger} \gamma^{\mu}\psi_j +(\psi \rightarrow \psi^c)
\label{2}
\ee
summed over fields, families and powers of $n,m$,
where $g'$ is the $U(1)'$ gauge coupling and we allow for different coupling strengths in the gauge coupling
denominator factors $\Lambda'$ as compared to $\Lambda$. The absence of a coupling at a given order 
corresponds to a particular $\Lambda$ or $\Lambda'$ becoming formally infinite.
In a given model, such as the example discussed in this paper, 
the various $\Lambda$ and $\Lambda'$ may be simply related.
The key feature of this scenario is that the same numerator factors 
of $\langle \phi_i \rangle$ control both the Yukawa couplings 
in Eq.\ref{1} and the $Z'$ couplings in Eq.\ref{2}.

Another key feature of the above scenario is that the $Z'$ mass is also generated by the VEVs $\langle \phi_i \rangle$,
so that $M_{Z'}\sim g'\langle \phi_i \rangle $.
This implies that the observation of $R_{K^*}$, which sets the scale of the $Z'$ mass and couplings, also sets the scale 
of the theory of flavour, which must both be not far from the TeV scale, $\langle \phi_i \rangle\sim {\rm TeV}$.
This does not happen in scalar leptoquark models, for example, since the scalar mass can be written
down by hand and it is not linked to the flavour scale.  Thus although the observation of $R_{K^*}$
fixes the mass of the mediator $Z'$ or leptoquark to be at the TeV scale,
it is only in the case of the $Z'$ that the VEV of $\langle \phi_i \rangle$ is necessarily also fixed to be at the TeV scale,
since its mass is given by $M_{Z'}\sim g'\langle \phi_i \rangle $.
By contrast, a TeV leptoquark is consistent with $\langle \phi_i \rangle$ 
and $\Lambda$ being much higher than the TeV scale, providing that their ratio is held fixed,
since its mass is not related to $\langle \phi_i \rangle$. For this reason we prefer to consider $Z'$ models.

In the scenario of Eqs.\ref{1},\ref{2},
in the limit that $\langle \phi_i \rangle =0$, there are no Yukawa couplings and also no 
couplings of SM fermions to the $Z'$ since we assume they are not charged under the associated $U(1)'$ gauge group. 
When $\langle \phi_i \rangle /\Lambda_i$
are switched on then Yukawa couplings and small non-universal and flavour dependent 
couplings of SM fermions to the $Z'$
are generated simultaneously, as well as the $Z'$ mass itself. 
The above framework then provides a link between flavour changing observables and 
the origin of Yukawa couplings of the kind that we are interested in.

In particular, there will be a connection between the experimental signal for new physics in 
$R_{K^*}$ due to $Z'$ exchange and the origin of Yukawa couplings.
Since the small Yukawa couplings are known (or at least their eigenvalues), 
this constrains the values of $\langle \phi_i \rangle /\Lambda_i$, and since we wish to explain $R_{K^*}$ via
non-universal $Z'$ exchange, then
this will also constrain the $Z'$ mass to be around the few TeV scale.
Then we have $M_{Z'}\sim \langle \phi_i \rangle \sim {\rm few \ TeV}$, and hence the theory of flavour is at the few TeV scale.
In general one might expect severe experimental flavour and collider constraints such an effective theory 
of flavour at the TeV scale. For example there could be dangerous flavour changing neutral currents (FCNCs) from
higher order operators with dimensional scale $\Lambda_i\sim  {\rm few \ TeV}$. However in such a scenario,
such FCNC operators are expected to have coefficients suppressed by small Yukawa couplings (generated by powers of 
$\langle \phi_i \rangle /\Lambda_i$). Of course this is not obvious from a general analysis, so one is motivated to 
provide some explicit renormalisable examples to substantiate this claim.

In fact there is a threefold motivation for going beyond the effective description in Eqs.\ref{1},\ref{2}.
Firstly, the treatment of the third family Yukawa couplings, especially the top quark coupling,
which must be considered carefully.
Secondly, for the Yukawa couplings, the scales $\langle \phi_i  \rangle $ and $\Lambda_i$
must be of order a few TeV, in order to explain $R_{K^*}$,
and so the effective theory in Eqs.\ref{1},\ref{2}
must break down not far above the TeV scale in any case.
Thirdly, in a concrete model
capable of describing all the quark and lepton masses and mixings, the relation between 
the Yukawa and $Z'$ couplings becomes very clear and it is more straightforward to see 
whether such a model can evade all the experimental constraints
while providing an explanation of 
$R_{K^*}$. In particular we can clearly investigate whether FCNC operators may be adequately suppressed
by small Yukawa as claimed above.
The main goal of this paper, then, is to show that the effective theory in Eqs.\ref{1},\ref{2}, with $Z'$ mass 
$M_{Z'}\sim g'\langle \phi_i \rangle $,
can be realised by some ultraviolet complete
renormalisable theory. 

Recently we have proposed 
an explanation of $R_{K^*}$ based on a fourth 
vector-like family which carries $U(1)'$, where the anomalies cancel between conjugate representations in the fourth family \cite{King:2017anf}. In the ``fermiophobic'' example, we showed that even if the three chiral families do not carry $U(1)'$ charges, 
effective non-universal couplings, as required to explain $R_{K^*}$, can emerge indirectly due to 
mixing with the fourth family \cite{King:2017anf}. 
A closely related idea was also considered some time later \cite{Raby:2017igl} in which the third chiral family and the conjugate 
half of the fourth vector-like family carried the $U(1)'$ charges, but otherwise the mechanism for $R_{K^*}$ works
very similar to \cite{King:2017anf}. Other phenomenological implications such as 
the muon $g-2$ and $\tau \rightarrow \mu \gamma$ were also considered \cite{Raby:2017igl}.
We have pursued this idea in several subsequent papers, including:
F-theory models with non-universal gauginos \cite{Romao:2017qnu};
$SO(10)$ models (including the question of neutrino mass) \cite{Antusch:2017tud};
$SU(5)$ models (focussing on the Yukawa relation $Y_e\neq Y_d^T$)  \cite{CarcamoHernandez:2018aon};
and flavourful $Z'$ portal models with a coupling to a fourth-family singlet Dirac neutrino Dark Matter,
including an extensive discussion of all phenomenological constraints
\cite{Falkowski:2018dsl}. In contrast to our previous work, here we shall require that small Yukawa couplings are 
not put in by hand, but are instead generated by couplings to the fourth family,
leading to the effective operators in Eqs.~\ref{1} and \ref{2}.

In this paper, then, we will construct an explicit renormalisable model with a vector-like family
as an ultraviolet completion of 
Eqs.~\ref{1} and \ref{2}. 
Our starting point is the model proposed in \cite{Ferretti:2006df},
which involves one vector-like family distinguished by a discrete $Z_2$.
We introduce a new $U(1)'$ gauge group, spontaneously broken at the TeV scale, under which the three chiral families 
are neutral but the vector-like fourth family is charged
\footnote{A nice consequence of this is that it is no longer necessary to introduce the discrete $Z_2$ symmetry.}.
The mixing between the fourth family and the three chiral families 
provides the Yukawa couplings as in Eq.\ref{1},
as well as the non-universal effective $Z'$ couplings involving the three light families
in Eq.\ref{2}, providing a link between the Yukawa and $Z'$ couplings.
However, while the fusion of these two ideas provides a nice connection between $R_{K^{(*)}}$ and the origin of Yukawa couplings in the quark sector, the lepton sector requires some tuning of Yukawa couplings to obtain 
the desired coupling of $Z'$ to muons.

The layout of the remainder of the paper is as follows.
In section~\ref{Anarchy} we consider the model  
with the vector-like family, where the Higgs is charged under the $U(1)'$
and effective Yukawa and $Z'$ couplings are generated by the same mixing.
Section~\ref{conclusion} concludes the paper.

\section{The model}
\label{Anarchy}
\subsection{The renormalisable Lagrangian}
The model we consider here is defined in Table~\ref{tab:funfields1}.
The model involves three chiral families $\psi_i(0), \psi^c_i(0)$, 
plus a fourth vector-like family consisting of $\psi_4(1), \psi^c_4(1)$ plus the conjugate representations
$\overline{\psi_4}(-1),\overline{\psi^c_4}(-1)$, where the $U(1)'$
charges are shown in parentheses.
The gauged $U(1)'$ is broken by the singlet scalars $\phi (1)$,
with vacuum expectation values (VEVs) around the TeV 
scale, yielding a massive $Z'$ at this scale. Since the Higgs doublets $H(-1)$
are charged under the $U(1)'$, this forbids all Yukawa couplings,
except those which couple the first three families to the fourth family.

A similar model was proposed as a model of effective Yukawa couplings in \cite{Ferretti:2006df}.
The main difference is that the model here involves a gauged $U(1)'$ 
resulting in effective Yukawa and flavourful $Z'$ couplings as in Eqs.\ref{1} and \ref{2}
which are related, while in \cite{Ferretti:2006df} only the effective Yukawa couplings were considered.
A welcome consequence of this is that, unlike \cite{Ferretti:2006df}, we shall not require an additional $Z_2$ symmetry to forbid 
renormalisable Yukawa couplings. Instead such couplings are forbidden by the gauged 
$U(1)'$ under which the fourth vector-like family and Higgs doublets are charged.
In addition, we shall go beyond the mass insertion approximation of \cite{Ferretti:2006df}, which breaks down for the top quark Yukawa coupling.
Another difference is that the model in \cite{Ferretti:2006df} was supersymmetric, while our model here is not.
However we require two Higgs doublets $H_u,H_d$, both with negative $U(1)'$ charge.

\begin{table}
\centering
\begin{tabular}{| l c c c c |}
\hline
Field & $SU(3)_c$ & $SU(2)_L$ & $U(1)_Y$ &$U(1)'$\\ 
\hline \hline
$Q_{i}$ 		 & ${\bf 3}$ & ${\bf 2}$ & $1/6$ & $0$ \\
$u^c_{i}$ 		 & ${\overline{\bf 3}}$ & ${\bf 1}$ & $-2/3$ & $0$\\
$d^c_{i}$ 		 & ${\overline{\bf 3}}$ & ${\bf 1}$ & $1/3$ & $0$\\
$L_{i}$ 		 & ${\bf 1}$ & ${\bf 2}$ & $-1/2$ & $0$\\
$e^c_{i}$ 		 & ${\bf 1}$ & ${\bf 1}$ & $1$ & $0$\\
$\nu^c_{i}$         & ${\bf 1}$ & ${\bf 1}$ & $0$ & $0$\\
\hline
\hline
$Q_{4}$ 		 & ${\bf 3}$ & ${\bf 2}$ & $1/6$ & $1$\\
$u^c_{4}$ 		 & ${\overline{\bf 3}}$ & ${\bf 1}$ & $-2/3$ & $1$\\
$d^c_{4}$ 		 & ${\overline{\bf 3}}$ & ${\bf 1}$ & $1/3$ & $1$\\
$L_{4}$ 		 & ${\bf 1}$ & ${\bf 2}$ & $-1/2$ & $1$\\
$e^c_{4}$ 		 & ${\bf 1}$ & ${\bf 1}$ & $1$ & $1$\\
$\nu^c_{4}$         & ${\bf 1}$ & ${\bf 1}$ & $0$ & $1$\\
\hline
\hline
$\overline{Q_{4}}$ 		 & $\overline{{\bf 3}}$ & $\overline{{\bf 2}}$ & $-1/6$ & $-1$\\
$\overline{u^c_{4}}$ 		 & ${{\bf 3}}$ & ${\bf 1}$ & $2/3$ & $-1$\\
$\overline{d^c_{4}}$ 		 & ${{\bf 3}}$ & ${\bf 1}$ & $-1/3$ & $-1$\\
$\overline{L_{4}}$ 		 & ${\bf 1}$ & $\overline{{\bf 2}}$ & $1/2$ & $-1$\\
$\overline{e^c_{4}}$ 		 & ${\bf 1}$ & ${\bf 1}$ & $-1$ & $-1$\\
$\overline{\nu^c_{4}}$         & ${\bf 1}$ & ${\bf 1}$ & $0$ & $-1$\\
\hline
\hline
$\phi$ & ${\bf 1}$ & ${\bf 1}$ & $0$ &$1$ \\
\hline
\hline
$H_u$ & ${\bf 1}$ & ${\bf 2}$ & $1/2$ &$-1$ \\
$H_d$ & ${\bf 1}$ & ${\bf 2}$ & $-1/2$ & $-1$\\
\hline
\end{tabular}
\caption{The model consists of 
three left-handed chiral families $\psi_i=Q_i,L_i$ and 
$\psi^c_i =u^c_i,d^c_i,e^c_i,\nu^c_i$ 
($i=1,2,3$),
plus a fourth vector-like family consisting of $\psi_4, \psi^c_4$ plus 
$\overline{\psi_4},\overline{\psi^c_4}$, together with the $U(1)'$ breaking scalar fields
$\phi$ and the two Higgs scalar doublets $H_u,H_d$ which are both negatively charged under $U(1)'$.}
\label{tab:funfields1}
\end{table}

\begin{figure}[ht]
\centering
	\includegraphics[scale=0.2]{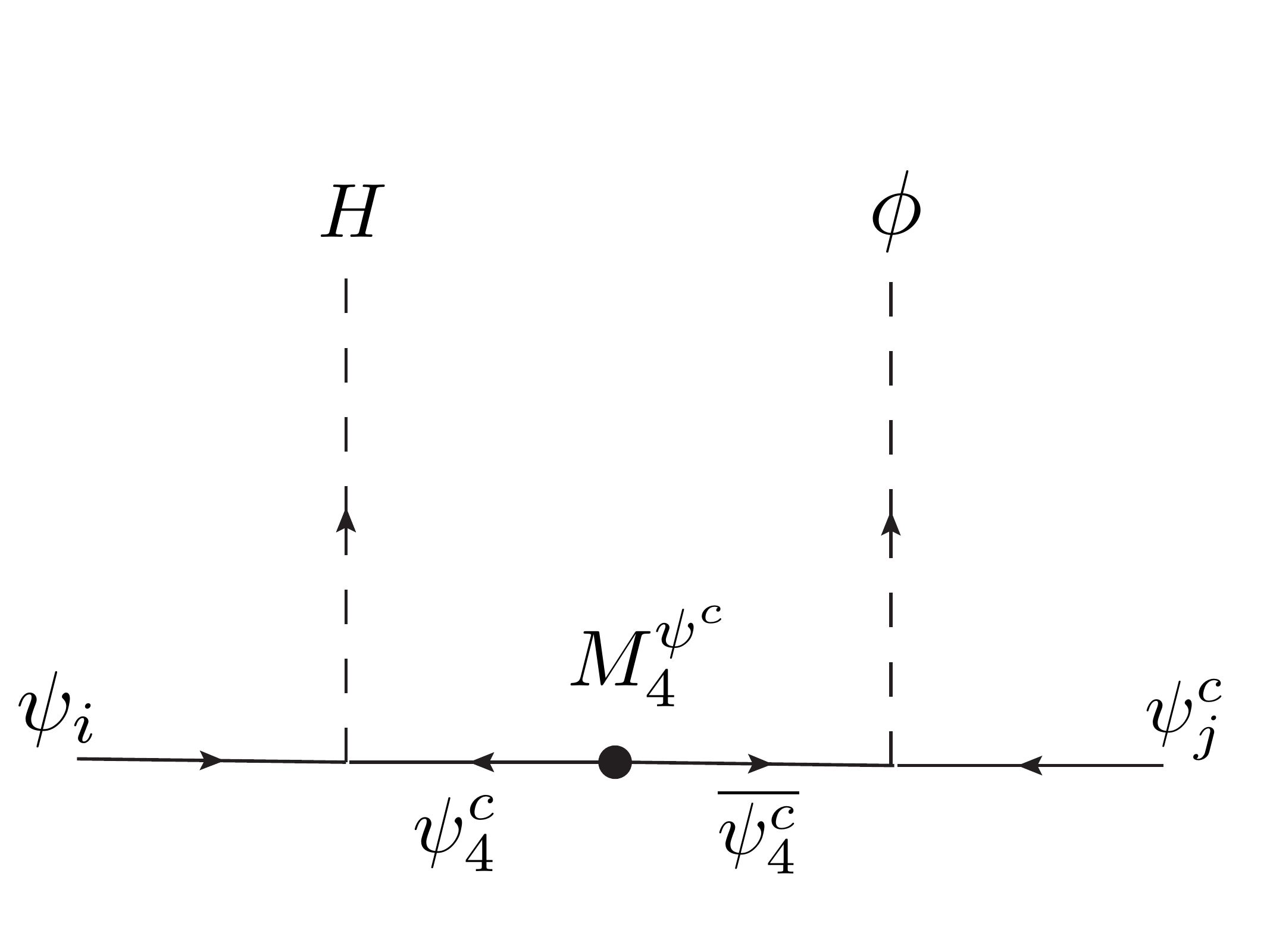}
\hspace*{1ex}
	\includegraphics[scale=0.2]{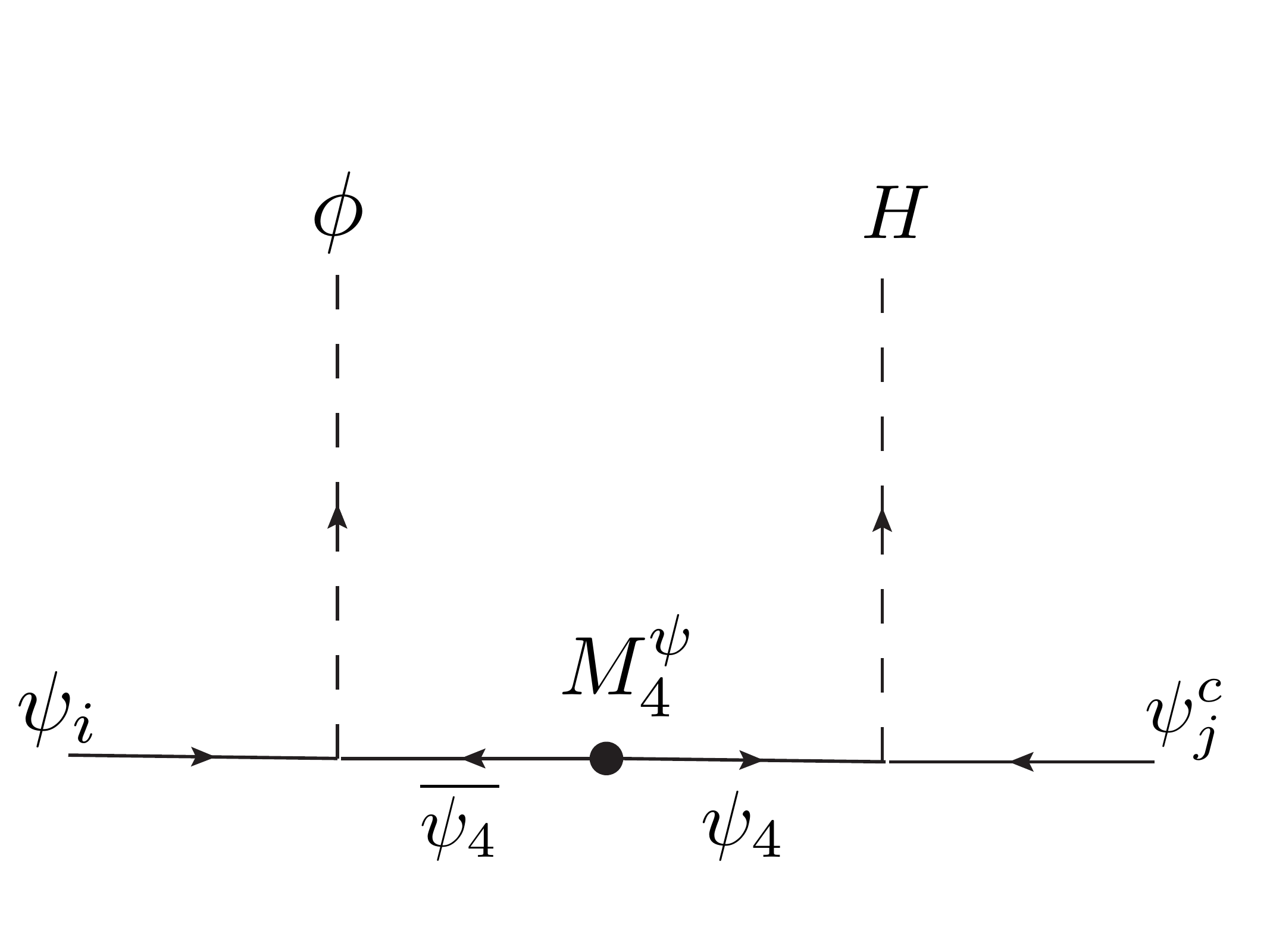}
\caption{Diagrams in the model which lead to the effective Yukawa couplings in the mass insertion approximation,
where $H=H_u,H_d$.}
\label{Fig1}
\end{figure}

\begin{figure}[ht]
\centering
	\includegraphics[scale=0.2]{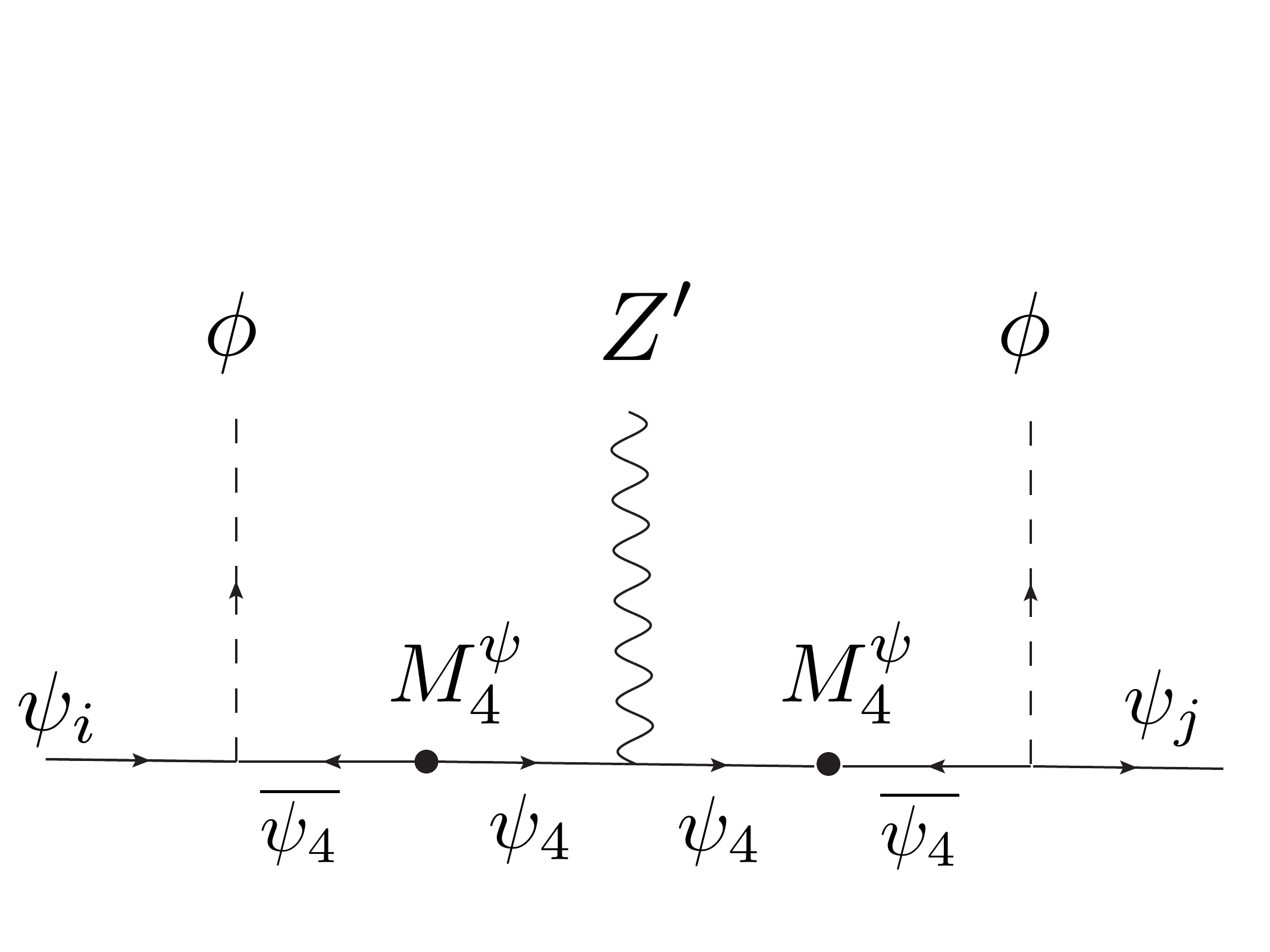}
\hspace*{1ex}
	\includegraphics[scale=0.2]{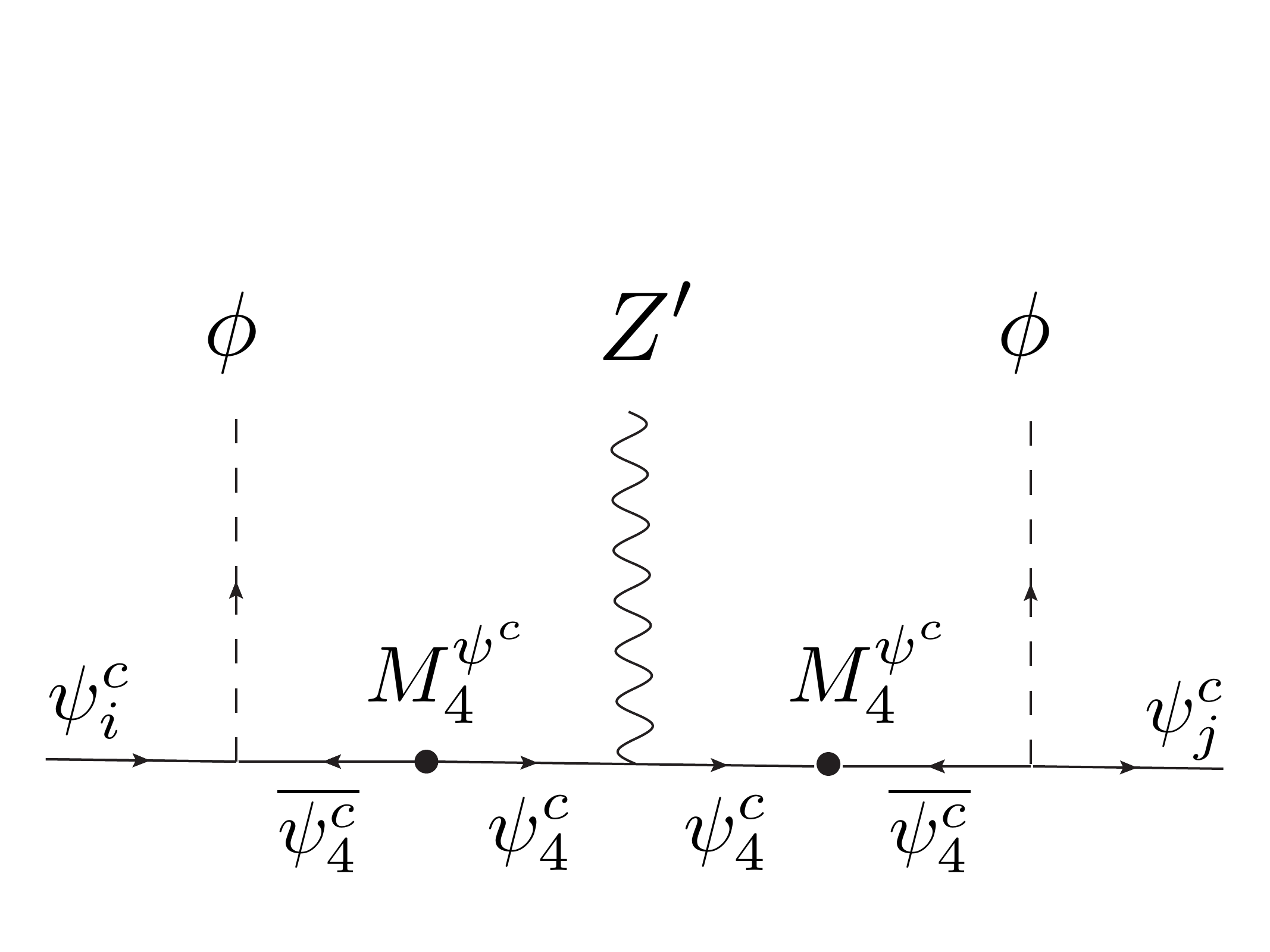}
\caption{Diagrams in the model which lead to the effective $Z'$ couplings in the mass insertion approximation.}
\label{Fig2}
\end{figure}

The allowed renormalisable Yukawa couplings and explicit masses allowed by $U(1)'$ are, 
\bea
{\cal L}^{ren} &=&
y^{\psi}_{i4}H  \psi_i {\psi^c_4} 
+  y^{\psi}_{4i}H {\psi_4} \psi^c_i+x^{\psi}_{i}\phi \psi_i \overline{\psi_4} 
+ x^{\psi^c}_{i}\phi \psi^c_i \overline{\psi^c_4}
+ M^{\psi}_{4}\psi_4 \overline{\psi_4}
+ M^{\psi^c}_{4}\psi^c_4 \overline{\psi^c_4}
\label{Anarchy_ren}
\eea
plus $H.c.$, summed over fields and families, where $x,y$ are dimensionless coupling constants ideally of order unity, while $M$ are explicit mass terms of order a few TeV. 

\subsection{Mass insertion approximation}

Although the usual Yukawa couplings $y^{\psi}_{ij} H  \psi_i {\psi^c_j}$ are forbidden for $i,j=1,\dots 3$
(since $H$ are charged under $U(1)'$) effective $3\times 3$ Yukawa couplings 
may be generated by the two mass insertion diagrams in Fig.\ref{Fig1} (up to an irrelevant minus sign),
\be
{\cal L}^{Yuk}_{eff}=\frac{x^{\psi^c}_{j} \langle \phi \rangle }{M^{\psi^c}_{4}} y^{\psi}_{i4} H  \psi_i {\psi^c_j}
+ \frac{x^{\psi}_{i} \langle \phi \rangle }{M^{\psi}_{4}} y^{\psi}_{4j}H  \psi_i {\psi^c_j}
\label{Yuk_mass_insertion}
\ee
plus $H.c.$, summed over fields and families, which can be compared to Eq.\ref{1}.

The model also involves a massive $Z'$
under which the three SM families $\psi_i,\psi_i^c$ have zero $U(1)'$ charge.
Although the usual $Z'$ couplings $g'Z'_{\mu}\psi_i^{\dagger} \gamma^{\mu}\psi_j $ are forbidden for $i,j=1,\dots 3$,
the fourth vector-like family has non-zero $U(1)'$ charge,
and effective $Z'$ couplings 
may be generated by the two mass insertion diagrams in Fig.\ref{Fig2},
\be
{\cal L}^{Z'}_{eff}=
\frac{x^{\psi}_{i}\langle \phi \rangle }{M^{\psi}_{4}}\frac{x^{\psi}_{j}\langle \phi \rangle }{M^{\psi}_{4}}
g'Z'_{\mu}\psi_i^{\dagger} \gamma^{\mu}\psi_j \ + \ 
\frac{x^{\psi^c}_{i}\langle \phi \rangle }{M^{\psi^c}_{4}} \frac{x^{\psi^c}_{j}\langle \phi \rangle }{M^{\psi^c}_{4}}
g'Z'_{\mu}\psi^{c \dagger}_i \gamma^{\mu}\psi^c_j 
\label{Zp_mass_insertion}
\ee
summed over fields and families, which can be compared to Eq.\ref{2}.
The above model is therefore an example of a renormalisable model which can lead
to the effective theory of the kind discussed in the Introduction, namely one in which 
Yukawa and $Z'$ couplings are both controlled by the same physics, in this case
the VEVs $\langle \phi \rangle$ and the fourth family vector-like masses 
$M^{\psi}_{4}$ and $M^{\psi^c}_{4}$. Moreover, the mass of the $Z'$ is given by 
$M_{Z'}=g'\langle \phi \rangle $, which is the same scale at which the Yukawa couplings are generated.
However, while the Yukawa couplings are generated at first order, the $Z'$ couplings are generated at second
order in the mass insertion approximation.
We shall discuss the phenomenological implications of this later.
For the moment, let us return to the Yukawa couplings and discuss them in some more detail.

There is a such a Yukawa matrix as in Eq.\ref{Yuk_mass_insertion} for each of the four charged sectors 
$\psi = u,d,e,\nu$. 
In the case of neutrinos, this refers to the Dirac Yukawa matrix, and there will be a further Majorana
mass matrix for the singlet neutrinos $M^{\nu^c}_{ij}\nu^c_i\nu^c_j$. Since nothing prevents the
Majorana masses $M^{\nu^c}_{ij}$
being arbitrarily large, well above the $U(1)'$ breaking scale, this will lead to a conventional seesaw mechanism
for small neutrino masses. On the other hand we are assuming that the vector-like masses 
$M^{\psi}_{4}$ and $M^{\psi^c}_{4}$ to be close to the $U(1)'$ breaking scale of order 
the TeV scale.

\subsection{The $5\times 5$ Matrix}

Since the large top quark Yukawa coupling $y_t$ is not present at renormalisable level, it must also arise from mixing with the fourth 
vector-like family. Clearly the mass insertion approximation in Eq.\ref{Yuk_mass_insertion} breaks down in the case
$y_t \sim y^u_{33}\sim 1$. 
The large top quark Yukawa coupling therefore motivates us to go beyond the mass insertion approximation used in 
\cite{Ferretti:2006df}. To proceed, we first arrange the masses and couplings in Eq.\ref{Anarchy_ren} into $5\times 5$ matrices,
one for each charge sector $\psi = u,d,e,\nu$,
\be
	M^{\psi} = \pmatr{
	&\psi^c_1&\psi^c_2&\psi^c_3&\psi^c_4&\overline{\psi_4}\\ 
	\hline
	\psi_1|&0&0&0&y^{\psi}_{14}H &x^{\psi}_{1}\phi\\
	\psi_2|&0&0&0&y^{\psi}_{24}H &x^{\psi}_{2}\phi\\
	\psi_3|&0&0&0&y^{\psi}_{34}H &x^{\psi}_{3}\phi\\ 
	\psi_4|& y^{\psi}_{41}H& y^{\psi}_{42}H& y^{\psi}_{43}H&0&M^{\psi}_{4}\\ 
	\overline{\psi^c_4}|&x^{\psi^c}_{1}\phi&x^{\psi^c}_{2}\phi&x^{\psi^c}_{3}\phi& M^{\psi^c}_{4}&0}.	
	\label{M^psi_an}
\ee
There are three distinct mass scales in these matrices: the Higgs VEVs $\langle H \rangle$, the $\phi$ VEVs $\langle \phi \rangle$
and the vector-like fourth family masses $M^{\psi}_{4}$, $M^{\psi^c}_{4}$. If all these mass scales are of the same order then 
the correct procedure is to diagonalise the full $5\times 5$ matrices in each of the charge sectors (apart from neutrinos which must be treated
differently due to the Majorana masses and the seesaw mechanism). Then unitarity violation will play a role.
However, in the approximation $\langle H \rangle \ll \langle \phi \rangle$ (physically $M_Z \ll M_{Z'}$),
it will not be necessary to diagonalise the full matrix in one step,
as we shall see.

\subsection{A convenient basis for quarks}
Since the upper $3\times 3$ block of Eq.\ref{M^psi_an} contains zeros we are free to rotate the first three families as we wish without changing the upper $3\times 3$ block. Similarly the $Z'$ couplings to the first three families remain zero under such rotations.
For example, in the quark sector we are allowed to 
go to a particular basis in $Q_i, d^c_i, u^c_i$ ($i=1,\ldots 3$) flavour space where 
$x^{Q}_{1,2}=0$, $y^{u}_{41,42}=0$, $y^{d}_{41,42}=0$.
Then we can further rotate the first and second families to set $x^{u^c}_1=0$, $x^{d^c}_1=0$ and $y^{u}_{14}=0$
(but in general $y^{d}_{14}\neq 0$ since the quark doublet rotations have already been used up).
In this basis the $5\times 5$ quark matrices $M^u,M^d$ from Eq.\ref{M^psi_an} become, respectively,
\be
	 \pmatr{
	&u^c_1& u^c_2&u^c_3&u^c_4&\overline{Q_4}\\ 
	\hline
	Q_1|&0&0&0&0 &0\\
	Q_2|&0&0&0&y^{u}_{24}H^u &0\\
	Q_3|&0&0&0&y^{u}_{34}H^u &x^{Q}_{3}\phi\\ 
	Q_4|& 0 & 0 & y^{u}_{43}H^u&0&M^{Q}_{4}\\ 
	\overline{u^c_4}|&0&x^{u^c}_{2}\phi&x^{u^c}_{3}\phi& M^{u^c}_{4}&0},
	 \pmatr{
	&d^c_1& d^c_2&d^c_3&d^c_4&\overline{Q_4}\\ 
	\hline
	Q_1|&0&0&0&y^{d}_{14}H^d &0\\
	Q_2|&0&0&0&y^{d}_{24}H^d &0\\
	Q_3|&0&0&0&y^{d}_{34}H^d &x^{Q}_{3}\phi\\ 
	Q_4|& 0 & 0 & y^{d}_{43}H^d&0&M^{Q}_{4}\\ 
	\overline{d^c_4}|&0&x^{d^c}_{2}\phi&x^{d^c}_{3}\phi& M^{d^c}_{4}&0}
	\label{M^quark_an_basis}
\ee
Note that the fifth column is identical for both matrices, but, after electroweak symmetry breaking, 
the quark doublets $Q$ are split apart into $u,d$ which become part of $M^u,M^d$, respectively.

For purposes of later comparison, it is instructive to show the result for the effective $3\times 3$ Yukawa matrices for the quarks, 
$y^{u}_{ij}H_uQ_iu^c_j$ and $y^{d}_{ij}H_dQ_id^c_j$,
obtained from the $5\times 5$ matrices in Eq.\ref{M^quark_an_basis} (in the above basis)
using the mass insertion approximation in Eq.\ref{Yuk_mass_insertion}, 
\bea
y^{u}_{ij}&=&  \pmatr{0& 0 & 0 \\
0 & y^{u}_{24} x^{u^c}_{2} & y^{u}_{24} x^{u^c}_{3}\\
0 & y^{u}_{34} x^{u^c}_{2} & y^{u}_{34} x^{u^c}_{3}
}
\frac{\langle \phi \rangle }{M^{u^c}_{4}}
+\pmatr{
0 & 0 & 0 \\
0 & 0 & 0 \\
0 & 0 & x^{Q}_{3} y^{u}_{43}} 
\frac{\langle \phi \rangle }{M^{Q}_{4}} \nonumber \\
y^{d}_{ij}&=&  \pmatr{
0 & y^{d}_{14} x^{d^c}_{2} & y^{d}_{14} x^{d^c}_{3} \\
0 & y^{d}_{24} x^{d^c}_{2} & y^{d}_{24} x^{d^c}_{3}\\
0 & y^{d}_{34} x^{d^c}_{2} & y^{d}_{34} x^{d^c}_{3}
}
\frac{\langle \phi \rangle }{M^{d^c}_{4}}
+\pmatr{
0 & 0 & 0 \\
0 & 0 & 0 \\
0 & 0 & x^{Q}_{3} y^{d}_{43}} 
\frac{\langle \phi \rangle }{M^{Q}_{4}}
\label{Yuk_mass_insertion_ud}
\eea
The effective Yukawa matrices above consist of the sum of two rank 1 matrices, so the first family will be massless without further modification.
The simplest modification is to include Higgs messengers as in \cite{Ferretti:2006df},
where such Higgs messengers are not charged under $U(1)'$ and so will not induce any
$Z'$ couplings to the first family.
Indeed the dominance of the fourth family diagrams over the Higgs exchange diagrams provides a nice explanation
of the smallness of the first family masses.
Here we shall not consider the Higgs messenger diagrams explicitly and instead work in the massless first family limit,
which is a good approximation and is sufficient to illustrate the connection with $Z'$ couplings which are only induced by
mixing with the fourth vector-like family.

If one of the two Yukawa terms in each of the expressions in Eq.\ref{Yuk_mass_insertion_ud} is dropped, then the 
second family masses will become zero as well. 
This suggests a natural explanation of the smallness of the second family compared to the first family, namely that 
one of the two terms dominates over the other one. 
This assumption was called ``messenger dominance'' in  \cite{Ferretti:2006df}. 
In order to account for the smallness of the CKM element $V_{cb}$ in the quark sector, it is natural to assume that 
the left-handed quark messengers dominate over the right-handed messengers, $M_4^Q\ll M_4^{d^c},M_4^{u^c}$,
which was called ``left-handed messenger dominance'' in  \cite{Ferretti:2006df},
with the further assumption $M_4^Q\ll M_4^{d^c}\ll M_4^{u^c}$ reproducing the more pronounced mass hierarchy
in the up sector than the down sector. Assuming all this leads to $|V_{cb}|\sim m_s/m_b$ 
with $V_{ub}$, though naturally small, being unconstrained  \cite{Ferretti:2006df}.
However to explain the smallness of the Cabibbo angle requires further model building 
such as an $SU(2)_R$ symmetry \cite{Ferretti:2006df}, although here we assume its smallness is accidental.

\subsection{A basis for decoupling the heavy fourth family}
As already remarked, given the large top Yukawa coupling, we need to go beyond mass insertion approximation
$ \langle \phi \rangle \ll M^{\psi}_{4}$, so we must return to the full $5\times 5$ mass matrices.
However we shall still assume that $\langle H \rangle \ll \langle \phi \rangle$ so that we may switch off the Higgs VEVs
all together in the first instance, and so obtain an effective SM after integrating out the heavy fourth family.
A first step in this direction is to work in a basis where the $5\times 5$ mass matrices have the form
\be
	M^{\psi} 
	= \pmatr{
	&\psi^c_1&\psi^c_2&\psi^c_3&\psi^c_4&\overline{\psi_4}\\ 
	\hline
	\psi_1|&&&&&0\\
	\psi_2|&&&& &0\\
	\psi_3|&&&\tilde{y}'^{\psi}_{\alpha \beta}H &&0\\ 
	\psi_4|&&&&&\tilde{M}^{\psi}_{4}\\ 
	\overline{\psi^c_4}|&0&0&0& \tilde{M}^{\psi^c}_{4}&0}.
	\label{M^psi_an_tilde}
\ee
where $\tilde{y}'^{\psi}_{\alpha \beta}$ with $\alpha, \beta = 1,\ldots 4$ are the $4\times 4$ 
upper block Yukawa matrices
in this basis.
The key feature of Eq.\ref{M^psi_an_tilde} are the zeros in the fifth row and column which are achieved by rotating the first four families
by the unitary $4\times 4$ transformations introduced in \cite{King:2017anf},
\be
V_{Q}=V^{Q}_{34}V^{Q}_{24}V^{Q}_{14}, \ \ V_{u^c}=V^{u^c}_{34}V^{u^c}_{24}V^{u^c}_{14}, 
 \ \ V_{d^c}=V^{d^c}_{34}V^{d^c}_{24}V^{d^c}_{14},
\label{VQparam}
\ee
where each of the unitary matrices $V_{i4}$ are parameterised by a single angle $\theta_{i4}$ describing the mixing between 
the $i$th chiral family and the $4$th vector-like family.

As emphasised in \cite{King:2017anf},
the basis in Eq.\ref{M^psi_an_tilde} is very useful for decoupling the heavy states since the first three rows and columns of the matrices do not involve any large mass terms. Then we can decouple the heavy fourth vector-like family, in the approximation 
$\langle H \rangle \ll \langle \phi \rangle $,
by simply striking out the fourth and fifth rows and columns of these matrices.
The Yukawa matrices of the SM, $\tilde{y}'^{\psi}_{ij}$, then correspond to the remaining $3\times 3 $ upper blocks of 
the $4\times 4 $ Yukawa matrices, $\tilde{y}'^{\psi}_{\alpha \beta}$.
However, it should be remembered that the undecoupled three families in this basis
contain admixtures of the original fourth vector-like family due to the mixing, and so will have modified Yukawa couplings,
as compared to the original (unmixed) three chiral families, as follows.

In the above basis, where only the fourth components of the fermions are very heavy,
the $4\times 4$ Yukawa couplings are,
\be
\tilde{y}'^u_{\alpha \beta} =V_{Q}\tilde{y}^u_{\alpha \beta} V_{u^c}^{\dagger}, \ \   
\tilde{y}'^d_{\alpha \beta} =V_{Q}\tilde{y}^d_{\alpha \beta} V_{d^c}^{\dagger}, 
\label{ytp}
\ee
and $\tilde{y}^u_{\alpha \beta},\tilde{y}^d_{\alpha \beta}$ are identified with the $4\times 4$ upper blocks of Eq.\ref{M^quark_an_basis}.
The effective SM Yukawa couplings for the quarks then
correspond to the $3\times 3 $ upper blocks of $\tilde{y}'^u_{\alpha \beta},\tilde{y}'^d_{\alpha \beta}$,
namely
\be
y^{u}_{ij}H_uQ_iu^c_j, \ \  y^{d}_{ij}H_dQ_id^c_j,\ \ {\rm with} \ \ y^{u}_{ij}\equiv \tilde{y}'^{u}_{ij}, 
\ \ y^{d}_{ij}\equiv \tilde{y}'^{d}_{ij}, \ \ (i,j=1,\ldots 3).
\label{SMYuks}
\ee
The effective SM Yukawa couplings have non-zero elements due to the mixing, even though originally they were all zero.
This is the origin of flavour in the low energy effective SM theory.

\subsection{Effective quark Yukawa couplings revisited}
 
We now calculate the effective SM Yukawa matrices for quarks in this particular basis, following the above 
more accurate procedure,
and then compare to the results to the mass insertion approximation in Eq.\ref{Yuk_mass_insertion_ud}.
This treatment still assumes that $\langle H \rangle \ll \langle \phi \rangle$ 
but relaxes the assumption $ \langle \phi \rangle \ll M^{Q}_{4}$, which is not valid
due to the large top quark Yukawa coupling. Therefore the treatment in this subsection is necessary
for consistency.

From Eq.\ref{M^quark_an_basis}  we read off the upper $4\times 4$ blocks,
\be
	\tilde{y}^u_{\alpha \beta}= \pmatr{
		0&0&0&0 \\
	0&0&0&y^{u}_{24} \\
	0&0&0&y^{u}_{34} \\ 
	 0 & 0 & y^{u}_{43}&0 },\quad
		 \tilde{y}^d_{\alpha \beta}=\pmatr{
		0&0&0&y^{d}_{14} \\
	0&0&0&y^{d}_{24} \\
	0&0&0&y^{d}_{34} \\ 
	 0 & 0 & y^{d}_{43}&0 }
	\label{yuyd_an_basis}
\ee
Eq.\ref{M^quark_an_basis} also shows that we need $(3,4)$ mixing in the $Q$ sector and 
both $(2,4)$ and $(3,4)$ mixing in the $u^c,d^c$ sectors to go to the decoupling basis in Eq.\ref{M^psi_an_tilde}.
The unitary matrices in Eq.\ref{VQparam} are then given by,
\be
	V_{Q}=V^{Q}_{34} = \pmatr{1&0&0&0\\0&1&0&0\\0&0&c^{Q}_{34}&s^{Q}_{34}
	\\ 0&0&-s^{Q}_{34}&c^{Q}_{34}}, \quad 
	s^{Q}_{34} = \frac{x^Q_3 \langle \phi \rangle }{ \sqrt{(x^Q_3 \langle \phi \rangle )^2+(M^{Q}_{4})^2 }},
		\label{34Q}
\ee
where we anticipate that $s^{Q}_{34} \sim 1$ due to the large top quark Yukawa coupling.
For $V_{u^c},V_{d^c}$ only the angles $\theta^{u^c}_{24}, \theta^{u^c}_{34}$ and 
$\theta^{d^c}_{24}, \theta^{d^c}_{34}$ may be non-zero, and indeed 
must be to generate the second family masses and the mixing angles (although
$ \theta^{u^c}_{34}$ is not necessary). 
However these angles are all small, since we assume
$M_4^Q\ll M_4^{d^c}\ll M_4^{u^c}$, so here we are allowed to approximate,
\be
	V_{u^c}=V^{u^c}_{34} V^{u^c}_{24}\approx \pmatr{1&0&0&0\\0&1&0& \theta^{u^c}_{24}\\0&0&1& \theta^{u^c}_{34}
	\\ 0&- \theta^{u^c}_{24}&- \theta^{u^c}_{34}&1}, \quad \theta^{u^c}_{24}\approx \frac{x_2^{u^c}\langle \phi \rangle }{M_4^{u^c}},
	\quad 
	 \theta^{u^c}_{34}\approx \frac{x_3^{u^c}\langle \phi \rangle }{M_4^{u^c}}
	 \label{3424uc}
	\ee
and similarly for $V_{d^c}$.

Given the results in Eqs.\ref{yuyd_an_basis},\ref{34Q},\ref{3424uc}, the $4\times 4$ matrices in the basis of Eq.\ref{ytp} may be readily
computed and the Yukawa couplings of the SM in Eq.\ref{SMYuks}
may then be identified as their upper $3\times 3$ blocks,
\bea
	{y}^u_{ij}&=& \pmatr{
		0&0&0 \\
	0& \theta^{u^c}_{24}y^{u}_{24}& \theta^{u^c}_{34} y^{u}_{24}\\
	0& c^{Q}_{34} \theta^{u^c}_{24}y^{u}_{34}& c^{Q}_{34} \theta^{u^c}_{34}y^{u}_{34}  }
	+\pmatr{
		0&0&0 \\
	0&0&0 \\
	0&0&s^{Q}_{34} y^{u}_{43} },\nonumber \\
		 {y}^d_{ij}&=&  \pmatr{
		0& \theta^{d^c}_{24}y^{d}_{14}& \theta^{d^c}_{34} y^{d}_{14}\\
	0& \theta^{d^c}_{24}y^{d}_{24}& \theta^{d^c}_{34} y^{d}_{24}\\
	0& c^{Q}_{34} \theta^{d^c}_{24}y^{d}_{34}& c^{Q}_{34} \theta^{d^c}_{34}y^{d}_{34}  }
		 +
	\pmatr{
		0&0&0 \\
	0&0&0 \\
	0&0&s^{Q}_{34} y^{d}_{43}   }
	\label{yuyd_exact}
\eea
The effective Yukawa matrices in Eq.\ref{yuyd_exact} reduce to those in the mass insertion approximation in Eq.\ref{Yuk_mass_insertion_ud} 
in the small $\theta^{Q}_{34}$ angle limit. However such an approximation is not justified since the top quark Yukawa coupling is identified
as $y_t \approx s^{Q}_{34} y^{u}_{43}$ which implies that $s^{Q}_{34} \sim 1$ as already anticipated.
We must use the more reliable expressions for the Yukawa matrices in Eq.\ref{yuyd_exact}.
The bottom quark Yukawa coupling $y_b \approx s^{Q}_{34} y^{d}_{43}$ may also be large if $\tan \beta = v_u/v_d$
is large.
The small values of the Yukawa coupling of the charm quark $y_c \approx  \theta^{u^c}_{24}y^{u}_{24}$
and strange quark $y_s \approx  \theta^{d^c}_{24}y^{d}_{24}$,
are accounted for by the 
assumption $M_4^Q\ll M_4^{d^c}\ll M_4^{u^c}$, which justifies the small angle approximations for the other angles.

\subsection{Effective $Z'$ couplings revisited}

As discussed in  \cite{King:2017anf}, there is an approximate 
GIM mechanism in the electroweak sector  involving $W^{\pm},Z$ boson exchange,
because all four families $\psi_{\alpha},\psi^c_{\alpha}$, transform identically under the SM gauge group,
so their mixing does not induce any such flavour violation.
However, this is not true for the $Z'$ gauge bosons, which only couple to the fourth family.
After the mixing with the fourth family, the three light families have induced non-universal and flavour violating couplings to the $Z'$,
which depend on the mixing angles of the $4\times 4$ unitary matrices in Eq.\ref{VQparam},
as we now discuss.

After $U(1)'$ breaking, we have a massive $Z'$ gauge boson which only couples to the fourth family 
in the basis of Eq.\ref{M^quark_an_basis}, ignoring the heavy neutrino singlets,
\be
{\cal L}^{gauge}_{Z'}= g'Z'_{\mu}\left(
{Q_{\alpha}^{\dagger}}D_Q\gamma^{\mu}{Q_{\beta}}
+ u_{\alpha}^{c \dagger}D_{u^c}\gamma^{\mu}u^c_{\beta}
+d_{\alpha}^{c \dagger}D_{d^c}\gamma^{\mu}d^c_{\beta}
+L_{\alpha}^{\dagger}D_L\gamma^{\mu}L_{\beta}
+e_{\alpha}^{c \dagger}D_{e^c}\gamma^{\mu}e^c_{\beta}
\right)
\label{gaugeZp}
\ee
where $\alpha, \beta = 1,\ldots 4$ and the diagonal $4\times 4$ charge matrices are
\be
D_Q=D_{u^c}=D_{d^c}=D_L=D_{e^c}={\rm diag}(0,0,0,1).
\label{Zpcharges}
\ee
Going to the basis of Eq.\ref{M^psi_an_tilde}, we have 
\be
{\cal L}^{gauge}_{Z'}= g'Z'_{\mu}\left(
{{Q'}_{\alpha}^{\dagger}}D'_Q\gamma^{\mu}{Q'_{\beta}}
+ {u'}_{\alpha}^{c \dagger}D'_{u^c}\gamma^{\mu}{u'}^c_{\beta}
+{d'}_{\alpha}^{c \dagger}D'_{d^c}\gamma^{\mu}{d'}^c_{\beta}
+{L'}_{\alpha}^{\dagger}D'_L\gamma^{\mu}L'_{\beta}
+{e'}_{\alpha}^{c \dagger}D'_{e^c}\gamma^{\mu}{e'}^c_{\beta}
\right)
\label{gaugeZpp}
\ee
where the rotated charge matrices are,
\be
D'_Q= V_{Q}D_QV_{Q}^{\dagger}, \ 
D'_{u^c}=V_{u^c}D_{u^c}V_{u^c}^{\dagger}, \ 
D'_{d^c}=V_{d^c}D_{d^c}V_{d^c}^{\dagger}
\label{Dp}
\ee
where $V_{Q},V_{u^c},V_{d^c}$ are given in Eqs.\ref{34Q},\ref{3424uc}.

Focussing on the upper $3\times 3$ blocks of $D'$,
the $Z'$ coupling matrices of the SM quarks are
\be
{\cal L}^{gauge}_{Z',3\times 3}={D'}^{Q}_{ij}g' Z'_{\mu}Q^{\dagger}_i\gamma^{\mu}Q_j+
{D'}^{u^c}_{ij}g' Z'_{\mu}u^{c \dagger}_i\gamma^{\mu}u^c_j+
{D'}^{d^c}_{ij}g' Z'_{\mu}d^{c \dagger}_i\gamma^{\mu}d^c_j,
\label{Zp_exact}
\ee
where the charge matrices may then be computed from the upper $3\times 3$ blocks of Eq.\ref{Dp},
\be
{D'}^Q_{ij}= \pmatr{
		0&0&0 \\
	0&0&0 \\
	0&0&(s^{Q}_{34})^2},\
{D'}^{u^c}_{ij}= \pmatr{
		0&0&0 \\
	0& (\theta^{u^c}_{24})^2& \theta^{u^c}_{24}\theta^{u^c}_{34} \\
	0& \theta^{u^c}_{24}\theta^{u^c}_{34} & (\theta^{u^c}_{34})^2  },\
{D'}^{d^c}_{ij}=   \pmatr{
		0&0&0 \\
	0& (\theta^{d^c}_{24})^2& \theta^{d^c}_{24}\theta^{d^c}_{34} \\
	0& \theta^{d^c}_{24}\theta^{d^c}_{34} & (\theta^{d^c}_{34})^2  }
\label{D_exact}
\ee
The important point is that the $Z'$ couplings of the SM quarks in Eqs.\ref{Zp_exact},\ref{D_exact}
are controlled by the same mixing angles that control their Yukawa couplings,
in the same basis Eq.\ref{yuyd_exact},
but are second order in these mixing angles. Thus while the large top quark Yukawa coupling implies that $s^{Q}_{34} \sim 1$ and hence
the $Z'$ couples in an unsuppressed way to the third family quark doublet $Q_3=(t_L,b_L)$,
there are no couplings to the first or second family quark doublets $Q_1=(u_L,d_L)$, $Q_2=(c_L,s_L)$ in the basis 
of Eq.\ref{yuyd_exact}.
Moreover, the small value of the Yukawa coupling of the charm quark $y_c \sim  \theta^{u^c}_{24}\sim m_c/m_t$
implies that the $c_R$ coupling to $Z'$ is suppressed by $(\theta^{u^c}_{24})^2\sim (m_c/m_t)^2\sim 10^{-4}$ in this basis.
The $s_R$ coupling to $Z'$ is similarly suppressed, so there there is a negligible contribution to $K_0-\bar{K_0}$ mixing
for $M_Z'\sim 1$ TeV.

\subsection{Phenomenology}
\label{pheno}

One way to explain the muon anomalies especially $R_{K^{(*)}}$ is via the couplings
\begin{equation}
{\cal L} \supset Z'_\mu \left(
g_{bs} \bar{s}_{L} \gamma^\mu b_{L} 
+ g_{\mu \mu}\bar{\mu}_{L} \gamma^\mu \mu_{L} 
\right),
\label{eq:Zp_Rk_couplings_desired_mixing}
\end{equation}
where in our model of quarks the above $Z'$ coupling originates from $g_{bb} \bar{b}_{L} \gamma^\mu b_{L} $
where $g_{bb}=g'(s^{Q}_{34})^2$ from Eq.\ref{D_exact}, where this coupling is 
in the basis where the quark Yukawa matrices are given by Eq.~\ref{Yuk_mass_insertion_ud}
(or equivalently Eq.~\ref{yuyd_exact}).
The CKM matrix for the quarks may be constructed in the usual way, by diagonalising these Yukawa matrices,
\be
V_{uL}y^uV^{\dagger}_{uR}= {\rm diag}(y_u,y_c,y_t), \ \ 
V_{dL}y^dV^{\dagger}_{dR}= {\rm diag}(y_d,y_s,y_b)
\label{diag}
\ee
to yield the unitary $3\times 3$ CKM matrix,
\be
V_{\rm CKM}=V_{uL}V^{\dagger}_{dL}.
\label{CKM}
\ee
The previous assumption of this model that $M_4^Q\ll M_4^{d^c}\ll M_4^{u^c}$ implies that the CKM mixing originates predominantly from the down sector, hence to good approximation,
\be
V_{\rm CKM}\approx V^{\dagger}_{dL}.
\label{CKM}
\ee
This implies that in the diagonal quark mass basis, the off-diagonal quark coupling in 
Eq.~\ref{eq:Zp_Rk_couplings_desired_mixing} is generated with 
\be
g_{bs}=g'(s^{Q}_{34})^2(V'^{\dagger}_{dL})_{32}\approx g'(s^{Q}_{34})^2 V_{ts}.
\label{gbs}
\ee
In our model, we expect $s^{Q}_{34}\sim 1/\sqrt{2}$, say, due to the large top Yukawa coupling,
with the gauge coupling $g'\sim 1$ and $V_{ts} \sim - 0.04$ (in the usual PDG convention for $V_{\rm CKM}$)
and so from Eq.\ref{gbs},
\be
g_{bs}=g'(s^{Q}_{34})^2 V_{ts} \sim - \frac{1}{50}.
\label{gbs2}
\ee
We have not yet specified the lepton sector so we do not yet know the value of $g_{\mu \mu}$.
In fact we will be guided by these anomalies in our construction of the lepton sector.
However we remark that $g_{\mu \mu}$ will have the same 
relative sign as  $g_{bb}$ (positive in our convention), hence the model predicts that 
$g_{\mu \mu}$ and $g_{bs}$ will
have the opposite relative sign as required to account for the $R_K$ and $R_{K^*}$ measurements~\cite{DiLuzio:2017fdq}.

As discussed in the Introduction, one possible explanation of the $R_K$ and $R_{K^*}$ measurements in LHCb is that the low-energy Lagrangian below the weak scale contains an additional contribution to the effective 4-fermion operator with left-handed muon, $b$-quark, and $s$-quark fields:  
\begin{equation}
\Delta \mathcal{L}_\text{eff} \supset G_{b s\mu} ( \bar{b}_L \gamma^\mu s_L) (\bar{\mu}_L \gamma_\mu \mu_L )+ {\rm h.c.}, \qquad G_{b s\mu} \approx \frac{1}{(31.5\text{ TeV})^2}. 
\label{eq:bsmu_fit}
\end{equation}
The positive sign of $G_{b s\mu}$ matches the required negative value of $C_9^{\mu}<0$ with 
$C_9^{\mu}= -C_{10}^{\mu}$.
The coefficient $G_{b s\mu}$ is due to $Z'$ exchange at tree-level and is given
as function of the couplings in Eq.~\ref{eq:Zp_Rk_couplings_desired_mixing},
\begin{equation}
G_{b s\mu} = -\frac{g_{bs} g_{\mu \mu}}{{M^2_{Z'}}}
\approx \frac{1}{(31.5\text{ TeV})^2}
\label{eq:bsmu_model}
\end{equation}
where the opposite relative sign of $g_{\mu \mu}$ and $g_{bs}$
is crucial to account for the $R_K$ and $R_{K^*}$ measurements~\cite{DiLuzio:2017fdq}.

The $Z^{\prime}$ coupling to $bs$ leads to an additional tree-level contribution to $B_s-\overline{B}_s$ mixing
due to the effective operator arising from $Z'$ exchange at tree level: 
\begin{equation}
\Delta \mathcal{L}_\text{eff} \supset  -\frac{g_{bs}^2 }{{2M^2_{Z'}}}  (\bar{s}_L \gamma^\mu b_L)^2 + {\rm h.c}, 
\label{eq:effective_bs}
\end{equation}
Such a new contribution is highly constrained by the measurements of the mass difference $\Delta M_s$ of neutral $B_s$ mesons 
as discussed in 2015 leading to the bound of $M_{Z'}/|g_{bs}|\gtrsim  150$ TeV \cite{Artuso:2015swg},
with a very recent 2017 bound from updated lattice results of 
$M_{Z'}/|g_{bs}|\gtrsim  500$ TeV \cite{DiLuzio:2017fdq}. 
However the stronger 2017 bound arises from a discrepancy with the Standard Model which could in principle disappear. 

If we take the milder $B_s-\overline{B}_s$ mixing bound then this constrains,
\begin{equation}
 \frac{|g_{bs}|}{{M_{Z'}}}
  \lesssim \frac{1}{150\text{ TeV}} .
\label{eq:coeff_Bs_only}
\end{equation} 
Since $g_{bs}$ is approximately known in our model, then using Eq.\ref{gbs}, this leads to a lower bound on the $Z'$ mass in this model,
\begin{equation}
 M_{Z'}
  \gtrsim 3\text{ TeV},
\label{Zpbound}
\end{equation} 
where the bound would increase by a factor of three for the stronger 2017 bound on $B_s-\overline{B}_s$ mixing. 

Since the Higgs doublets are charged under $U(1)'$, they will induce $Z-Z'$ mixing which will affect the
SM prediction of $M_W/M_Z$, leading to corrections to the well determined 
electroweak precision parameter $\rho$ or $T$. One may estimate the lower bound\footnote{Adam Falkowski,
private communication.},
\begin{equation}
 \frac{M_{Z'}}{g'k}
  \gtrsim 9.5\text{ TeV},
\label{Zpbound2}
\end{equation} 
where $k$ is the model dependent normalisation of the $U(1)'$ charge of the Higgs doublets.

Taking the ratio of Eqs. \ref{eq:coeff_Bs_only} to \ref{eq:bsmu_model}
we find the relatively model independent bound,
\begin{equation}
\frac{|g_{bs}|}{g_{\mu \mu}} \lesssim \frac{1}{25}.
\label{ratio}
\end{equation} 
However this would reduce by an order of magnitude for the stronger 2017 bound on $B_s-\overline{B}_s$ mixing. 
Since $g_{bs}$ is approximately known in our model, then Eqs.\ref{ratio} and \ref{gbs} imply the bound 
\begin{equation}
g_{\mu \mu}\gtrsim \frac{1}{2}.
\label{gmumubound}
\end{equation} 
If the $U(1)'$ gauge charges corresponded to a fourth family $B-L$ then this bound would become $g_{\mu \mu}\gtrsim 1/6$
but the coupling $g_{\mu \mu}$ must still be quite sizeable in any case.
The message is clear that in order to account for $R_{K^{(*)}}$ we need to generate an unsuppressed coupling of $Z'$
to muons. In particular we need a large value of $\theta_{24}^L$, as observed in \cite{Falkowski:2018dsl}.
Assuming this, then we satisfy the phenomenological constraints from $Z'$ searches, $B_s-\overline{B}_s$ mixing,
trident events and so on
as fully discussed in \cite{Falkowski:2018dsl}
to which we refer the reader for more details.
We now return to the question of how it is possible to obtain large muon couplings to $Z'$ in this model,
and show that this may only be achieved at the expense of tuning of Yukawa couplings.

\subsection{Leptons}
\label{leptons}
The leptons can be treated by exactly the same methods as developed for the quarks.
However, the situation for the leptons is slightly different since there will be a further Majorana
mass matrix for the singlet neutrinos $M^{\nu^c}_{ij}\nu^c_i\nu^c_j$. Since nothing prevents the
Majorana masses $M^{\nu^c}_{ij}$
being arbitrarily large, well above the $U(1)'$ breaking scale, this will lead to a conventional seesaw mechanism
for small neutrino masses. In principle, large lepton mixing can originate from 
the seesaw mechanism due to the arbitrary neutrino Yukawa matrix and the heavy Majorana mass matrix 
$M^{\nu^c}_{ij}$, so we are free to treat the charged lepton Yukawa matrices any way we wish.

Motivated by the $R_{K^{(*)}}$ anomalies, we work in a new basis for the charged leptons,
by performing suitable rotations on the fields $L_i, e^c_i$, to give, 
\be
	 \pmatr{
	&e^c_1& e^c_2&e^c_3&e^c_4&\overline{L_4}\\ 
	\hline
	L_1|&0&0&0&0 &0\\
	L_2|&0&0&0&y^{e}_{24}H^d &x^{L}_{2}\phi\\
	L_3|&0&0&0&y^{e}_{34}H^d &0\\ 
	L_4|& 0 & y^{e}_{42}H^d & 0 &0&M^{L}_{4}\\ 
	\overline{e^c_4}|&0&x^{e^c}_{2}\phi&x^{e^c}_{3}\phi& M^{e^c}_{4}&0}.
	 	\label{M^e_new_basis}
\ee

Now assuming all angles to be small apart from $\theta_{24}^L$, motivated by the discussion in the previous subsection~\cite{Falkowski:2018dsl},
we obtain, 
\be
	{y}^e_{ij}= \pmatr{
		0& 0&0\\
	0&c^L_{24}  \theta^{e^c}_{24}y^{e}_{24}& c^L_{24} \theta^{e^c}_{34} y^{e}_{24}\\
	0& \theta^{e^c}_{24}y^{e}_{34}&   \theta^{e^c}_{34}y^{e}_{34}  }
	+\pmatr{
		0&0&0 \\
	0&s^{L}_{24} y^{e}_{42} &0 \\
	0&0&0 },
	\label{ye_exact}
\ee
The $Z'$ coupling matrices in this basis (ignoring the heavy Majorana neutrinos) are,
\be
{D}^L_{ij}= \pmatr{
		0&0&0 \\
	0&(s^{L}_{24})^2&0 \\
	0&0&0},\
{D}^{e^c}_{ij}= \pmatr{
		0&0&0 \\
	0& (\theta^{e^c}_{24})^2& \theta^{e^c}_{24}\theta^{e^c}_{34} \\
	0& \theta^{e^c}_{24}\theta^{e^c}_{34} & (\theta^{e^c}_{34})^2  }.\
\label{De_exact}
\ee
So far this looks very similar to the basis used for the up type quarks, in Eqs.\ref{yuyd_exact}, \ref{D_exact}
with the replacements $Q\rightarrow L$, with $u^c\rightarrow e^c$ and
a relabelling $L_2\leftrightarrow L_3$ and $e^c_2\leftrightarrow e^c_3$.
Indeed, without further assumption, the Yukawa matrices seem to be dominated by the element with the largest angle,
which would imply that the second family charged lepton is the heaviest, so we would interpret that as the $\tau$ lepton.
However, let us suppose that we tune the Yukawa coupling $y^{e}_{42}$ to be very small in this basis,
with the hierarchy
$y^{e}_{42}\ll y^{e}_{34}$ so that the charged lepton Yukawa matrix is in fact dominated by the
first matrix in Eq.\ref{ye_exact}, even though the angles are assumed to be small. 
This may also lead to large left-handed charged lepton mixing,
providing an explanation of large atmospheric lepton mixing in neutrino oscillations.
Moreover, since the first matrix is rank one,
the smaller muon mass is controlled by $y^{e}_{42}$ from the second matrix in Eq.\ref{ye_exact}. 
\footnote{Alternatively, we may also tune
$y^{e}_{24}\ll y^{e}_{34}$ so there is small left-handed charged lepton mixing. 
Most of the large lepton mixing would then originate in the neutrino sector.}
We conclude that this explanation of $R_{K^{(*)}}$ requires the Yukawa coupling $y^{e}_{42}$ to be tuned to be small.

\section{Conclusions}
\label{conclusion}

We have explored the possibility that the semi-leptonic $B$ decay ratios $R_{K^{(*)}}$ which violate $\mu - e$ universality
are related to the origin of the fermion Yukawa couplings in the Standard Model. 
We have constructed an explicit renormalisable model with a vector-like family
as an ultraviolet completion of 
Eqs.~\ref{1} and \ref{2}. 
The considered model involves 
a new $U(1)'$ gauge group, spontaneously broken at the TeV scale, under which the three chiral families 
are neutral but the vector-like fourth family is charged.
The mixing between the fourth family and the three chiral families 
then provides the effective Yukawa couplings, as in Fig.~\ref{Fig1},
as well as the non-universal effective $Z'$ couplings involving the three light families,
in Fig.~\ref{Fig2}, which are both generated by the same physics.
However, while the model provides a nice connection between $R_{K^{(*)}}$ and the origin of Yukawa couplings in the quark sector, the lepton sector requires some tuning of Yukawa couplings to obtain 
the desired coupling of $Z'$ to muons.

\subsection*{Acknowledgements}

S.\,F.\,K. acknowledges the STFC Consolidated Grant ST/L000296/1 and the European Union's Horizon 2020 Research and Innovation programme under Marie Sk\l{}odowska-Curie grant agreements Elusives ITN No.\ 674896 and InvisiblesPlus RISE No.\ 690575.


\begin{thebibliography}{99}
 \setlength{\itemsep}{0em}
 
\bibitem{Langacker:2008yv}
  P.~Langacker,
  Rev.\ Mod.\ Phys.\  {\bf 81} (2009) 1199
  doi:10.1103/RevModPhys.81.1199
  [arXiv:0801.1345 [hep-ph]].
    
  



  
\bibitem{Descotes-Genon:2013wba} 
  S.~Descotes-Genon, J.~Matias and J.~Virto,
  Phys.\ Rev.\ D {\bf 88}, 074002 (2013)
  doi:10.1103/PhysRevD.88.074002
  [arXiv:1307.5683 [hep-ph]].



\bibitem{Altmannshofer:2013foa} 
  W.~Altmannshofer and D.~M.~Straub,
  Eur.\ Phys.\ J.\ C {\bf 73}, 2646 (2013)
  doi:10.1140/epjc/s10052-013-2646-9
  [arXiv:1308.1501 [hep-ph]].



\bibitem{Ghosh:2014awa} 
  D.~Ghosh, M.~Nardecchia and S.~A.~Renner,
  JHEP {\bf 1412}, 131 (2014)
  doi:10.1007/JHEP12(2014)131
  [arXiv:1408.4097 [hep-ph]].



\bibitem{Aaij:2014ora} 
  R.~Aaij {\it et al.} [LHCb Collaboration],
  Phys.\ Rev.\ Lett.\  {\bf 113}, 151601 (2014)
  doi:10.1103/PhysRevLett.113.151601
  [arXiv:1406.6482 [hep-ex]].


 \bibitem{Bifani}
S.~Bifani for the LHCb Collaboration, {\it Search for new physics with $b \to s \ell^+ \ell^-$ decays at LHCb}, CERN Seminar, 18 April 2017,
{\tt https://cds.cern.ch/record/2260258}.


\bibitem{Hiller:2017bzc} 
  G.~Hiller and I.~Nisandzic,
  Phys.\ Rev.\ D {\bf 96}, no. 3, 035003 (2017)
  doi:10.1103/PhysRevD.96.035003
  [arXiv:1704.05444 [hep-ph]].


\bibitem{Ciuchini:2017mik}
  M.~Ciuchini, A.~M.~Coutinho, M.~Fedele, E.~Franco, A.~Paul, L.~Silvestrini and M.~Valli,
  Eur.\ Phys.\ J.\ C {\bf 77} (2017) no.10,  688
  doi:10.1140/epjc/s10052-017-5270-2
  [arXiv:1704.05447 [hep-ph]].



\bibitem{Geng:2017svp} 
  L.~S.~Geng, B.~Grinstein, S.~Jäger, J.~Martin Camalich, X.~L.~Ren and R.~X.~Shi,
  Phys.\ Rev.\ D {\bf 96}, no. 9, 093006 (2017)
  doi:10.1103/PhysRevD.96.093006
  [arXiv:1704.05446 [hep-ph]].



\bibitem{Capdevila:2017bsm} 
  B.~Capdevila, A.~Crivellin, S.~Descotes-Genon, J.~Matias and J.~Virto,
  JHEP {\bf 1801}, 093 (2018)
  doi:10.1007/JHEP01(2018)093
  [arXiv:1704.05340 [hep-ph]].



\bibitem{Ghosh:2017ber} 
  D.~Ghosh,
  Eur.\ Phys.\ J.\ C {\bf 77}, no. 10, 694 (2017)
  doi:10.1140/epjc/s10052-017-5282-y
  [arXiv:1704.06240 [hep-ph]].



\bibitem{Bardhan:2017xcc} 
  D.~Bardhan, P.~Byakti and D.~Ghosh,
  Phys.\ Lett.\ B {\bf 773}, 505 (2017)
  doi:10.1016/j.physletb.2017.08.062
  [arXiv:1705.09305 [hep-ph]].



\bibitem{Glashow:2014iga} 
  S.~L.~Glashow, D.~Guadagnoli and K.~Lane,
  Phys.\ Rev.\ Lett.\  {\bf 114}, 091801 (2015)
  doi:10.1103/PhysRevLett.114.091801
  [arXiv:1411.0565 [hep-ph]].



\bibitem{DAmico:2017mtc} 
  G.~D'Amico, M.~Nardecchia, P.~Panci, F.~Sannino, A.~Strumia, R.~Torre and A.~Urbano,
  JHEP {\bf 1709}, 010 (2017)
  doi:10.1007/JHEP09(2017)010
  [arXiv:1704.05438 [hep-ph]].



\bibitem{Descotes-Genon:2015uva}
  S.~Descotes-Genon, L.~Hofer, J.~Matias and J.~Virto,
  JHEP {\bf 1606} (2016) 092
  doi:10.1007/JHEP06(2016)092
  [arXiv:1510.04239 [hep-ph]].


\bibitem{Calibbi:2015kma}
  L.~Calibbi, A.~Crivellin and T.~Ota,
  Phys.\ Rev.\ Lett.\  {\bf 115} (2015) 181801
  doi:10.1103/PhysRevLett.115.181801
  [arXiv:1506.02661 [hep-ph]].
  
  
  
  
  

  \bibitem{large}
 R.~Gauld, F.~Goertz and U.~Haisch,
  JHEP {\bf 1401} (2014) 069
  doi:10.1007/JHEP01(2014)069
  [arXiv:1310.1082 [hep-ph]];
  A.~J.~Buras and J.~Girrbach,
  JHEP {\bf 1312} (2013) 009
  doi:10.1007/JHEP12(2013)009
  [arXiv:1309.2466 [hep-ph]];
 A.~J.~Buras, F.~De Fazio and J.~Girrbach,
  JHEP {\bf 1402} (2014) 112
  doi:10.1007/JHEP02(2014)112
  [arXiv:1311.6729 [hep-ph]];
  W.~Altmannshofer, S.~Gori, M.~Pospelov and I.~Yavin,
  Phys.\ Rev.\ D {\bf 89} (2014) 095033
  doi:10.1103/PhysRevD.89.095033
  [arXiv:1403.1269 [hep-ph]];
  A.~Crivellin, G.~D'Ambrosio and J.~Heeck,
  Phys.\ Rev.\ Lett.\  {\bf 114} (2015) 151801
  doi:10.1103/PhysRevLett.114.151801
  [arXiv:1501.00993 [hep-ph]]
and  
  Phys.\ Rev.\ D {\bf 91} (2015) no.7,  075006
  doi:10.1103/PhysRevD.91.075006
  [arXiv:1503.03477 [hep-ph]];
  C.~Niehoff, P.~Stangl and D.~M.~Straub,
  Phys.\ Lett.\ B {\bf 747} (2015) 182
  doi:10.1016/j.physletb.2015.05.063
  [arXiv:1503.03865 [hep-ph]];
  A.~Celis, J.~Fuentes-Martin, M.~Jung and H.~Serodio,
  Phys.\ Rev.\ D {\bf 92} (2015) no.1,  015007
  doi:10.1103/PhysRevD.92.015007
  [arXiv:1505.03079 [hep-ph]];
 A.~Greljo, G.~Isidori and D.~Marzocca,
  JHEP {\bf 1507} (2015) 142
  doi:10.1007/JHEP07(2015)142
  [arXiv:1506.01705 [hep-ph]];
   W.~Altmannshofer and I.~Yavin,
  Phys.\ Rev.\ D {\bf 92} (2015) no.7,  075022
  doi:10.1103/PhysRevD.92.075022
  [arXiv:1508.07009 [hep-ph]];
 A.~Falkowski, M.~Nardecchia and R.~Ziegler,
  JHEP {\bf 1511} (2015) 173
  doi:10.1007/JHEP11(2015)173
  [arXiv:1509.01249 [hep-ph]];
  B.~Allanach, F.~S.~Queiroz, A.~Strumia and S.~Sun,
  Phys.\ Rev.\ D {\bf 93} (2016) no.5,  055045
  doi:10.1103/PhysRevD.93.055045
  [arXiv:1511.07447 [hep-ph]];
   C.~W.~Chiang, X.~G.~He and G.~Valencia,
  Phys.\ Rev.\ D {\bf 93} (2016) no.7,  074003
  doi:10.1103/PhysRevD.93.074003
  [arXiv:1601.07328 [hep-ph]];
  S.~M.~Boucenna, A.~Celis, J.~Fuentes-Martin, A.~Vicente and J.~Virto,
  Phys.\ Lett.\ B {\bf 760} (2016) 214
  doi:10.1016/j.physletb.2016.06.067
  [arXiv:1604.03088 [hep-ph]]
and
  JHEP {\bf 1612} (2016) 059
  doi:10.1007/JHEP12(2016)059
  [arXiv:1608.01349 [hep-ph]];
  P.~Ko, Y.~Omura, Y.~Shigekami and C.~Yu,
  arXiv:1702.08666 [hep-ph].
  K.~Ishiwata, Z.~Ligeti and M.~B.~Wise,
  JHEP {\bf 1510} (2015) 027
  doi:10.1007/JHEP10(2015)027
  [arXiv:1506.03484 [hep-ph]];
  D.~Aristizabal Sierra, F.~Staub and A.~Vicente,
  Phys.\ Rev.\ D {\bf 92} (2015) no.1,  015001
  doi:10.1103/PhysRevD.92.015001
  [arXiv:1503.06077 [hep-ph]];
  G.~Belanger, C.~Delaunay and S.~Westhoff,
  Phys.\ Rev.\ D {\bf 92} (2015) 055021
  doi:10.1103/PhysRevD.92.055021
  [arXiv:1507.06660 [hep-ph]];
  C.~Bobeth, A.~J.~Buras, A.~Celis and M.~Jung,
  JHEP {\bf 1704} (2017) 079
  doi:10.1007/JHEP04(2017)079
  [arXiv:1609.04783 [hep-ph]].



 \bibitem{recent}
  C.~W.~Chiang, X.~G.~He, J.~Tandean and X.~B.~Yuan,
  arXiv:1706.02696 [hep-ph];
  Y.~Tang and Y.~L.~Wu,
  arXiv:1705.05643 [hep-ph];
  F.~Bishara, U.~Haisch and P.~F.~Monni,
  arXiv:1705.03465 [hep-ph];
  J.~Ellis, M.~Fairbairn and P.~Tunney,
  arXiv:1705.03447 [hep-ph];
  J.~F.~Kamenik, Y.~Soreq and J.~Zupan,
  arXiv:1704.06005 [hep-ph];
  F.~Sala and D.~M.~Straub,
  arXiv:1704.06188 [hep-ph];
  S.~Di Chiara, A.~Fowlie, S.~Fraser, C.~Marzo, L.~Marzola, M.~Raidal and C.~Spethmann,
  arXiv:1704.06200 [hep-ph];
  A.~K.~Alok, B.~Bhattacharya, A.~Datta, D.~Kumar, J.~Kumar and D.~London,
  arXiv:1704.07397 [hep-ph];
  R.~Alonso, P.~Cox, C.~Han and T.~T.~Yanagida,
  arXiv:1704.08158 [hep-ph];
  C.~Bonilla, T.~Modak, R.~Srivastava and J.~W.~F.~Valle,
  arXiv:1705.00915 [hep-ph];
  S.~F.~King,
  JHEP {\bf 1708} (2017) 019
  doi:10.1007/JHEP08(2017)019
  [arXiv:1706.06100 [hep-ph]].


\bibitem{Aloni:2017ixa}
  D.~Aloni, A.~Dery, C.~Frugiuele and Y.~Nir,
  JHEP {\bf 1711} (2017) 109
  doi:10.1007/JHEP11(2017)109
  [arXiv:1708.06161 [hep-ph]];
  S.~Y.~Guo, Z.~L.~Han, B.~Li, Y.~Liao and X.~D.~Ma,
  Nucl.\ Phys.\ B {\bf 928} (2018) 435
  doi:10.1016/j.nuclphysb.2018.01.024
  [arXiv:1707.00522 [hep-ph]];
  G.~Hiller and I.~Nisandzic,
  Phys.\ Rev.\ D {\bf 96} (2017) no.3,  035003
  doi:10.1103/PhysRevD.96.035003
  [arXiv:1704.05444 [hep-ph]];
  H.~Päs and E.~Schumacher,
  Phys.\ Rev.\ D {\bf 92} (2015) no.11,  114025
  doi:10.1103/PhysRevD.92.114025
  [arXiv:1510.08757 [hep-ph]];
  I.~de Medeiros Varzielas and G.~Hiller,
  JHEP {\bf 1506} (2015) 072
  doi:10.1007/JHEP06(2015)072
  [arXiv:1503.01084 [hep-ph]].




\bibitem{King:2017anf} 
  S.~F.~King,
  JHEP {\bf 1708}, 019 (2017)
  doi:10.1007/JHEP08(2017)019
  [arXiv:1706.06100 [hep-ph]].

\bibitem{Raby:2017igl}
  S.~Raby and A.~Trautner,
  Phys.\ Rev.\ D {\bf 97} (2018) no.9,  095006
  doi:10.1103/PhysRevD.97.095006
  [arXiv:1712.09360 [hep-ph]].

\bibitem{Romao:2017qnu} 
  M.~C.~Romao, S.~F.~King and G.~K.~Leontaris,
  arXiv:1710.02349 [hep-ph].



\bibitem{Antusch:2017tud} 
  S.~Antusch, C.~Hohl, S.~F.~King and V.~Susic,
  arXiv:1712.05366 [hep-ph].
  
\bibitem{CarcamoHernandez:2018aon}
  A.~E.~Carcamo Hernandez and S.~F.~King,
  arXiv:1803.07367 [hep-ph].

  
\bibitem{Falkowski:2018dsl}
  A.~Falkowski, S.~F.~King, E.~Perdomo and M.~Pierre,
  arXiv:1803.04430 [hep-ph].
  
  
\bibitem{Ferretti:2006df}
  L.~Ferretti, S.~F.~King and A.~Romanino,
  JHEP {\bf 0611} (2006) 078
  doi:10.1088/1126-6708/2006/11/078
  [hep-ph/0609047];
  L.~Calibbi, L.~Ferretti, A.~Romanino and R.~Ziegler,
  JHEP {\bf 0903} (2009) 031
  doi:10.1088/1126-6708/2009/03/031
  [arXiv:0812.0087 [hep-ph]].
  
  
\bibitem{DiLuzio:2017fdq}
  L.~Di Luzio, M.~Kirk and A.~Lenz,
  Phys.\ Rev.\ D {\bf 97} (2018) no.9,  095035
  doi:10.1103/PhysRevD.97.095035
  [arXiv:1712.06572 [hep-ph]].

  
\bibitem{Artuso:2015swg}
  M.~Artuso, G.~Borissov and A.~Lenz,
  Rev.\ Mod.\ Phys.\  {\bf 88} (2016) no.4,  045002
  doi:10.1103/RevModPhys.88.045002
  [arXiv:1511.09466 [hep-ph]].
  
  
    
  
\bibitem{Nir:2016zkd}
  Y.~Nir,
  CERN-2015-001, pp.123-156
  doi:10.5170/CERN-2015-001.123
  [arXiv:1605.00433 [hep-ph]].

  
  
  
  
 
  
  
    
  
  
  

\end{thebibliography}
\end{document}